\RequirePackage{fix-cm}
\documentclass[twocolumn,epjc3]{svjour3}  

\usepackage{cite}
\usepackage{amsmath}
\usepackage{amssymb}
\usepackage{amsfonts}
\usepackage{appendix}
\usepackage{orcidlink}

\usepackage{lipsum}     
\usepackage{mathtools}
\usepackage{cuted}
\usepackage{graphics}
\usepackage{latexsym}
\usepackage{orcidlink}
\usepackage{enumerate}

\smartqed  % flush right qed marks, e.g. at end of proof
\RequirePackage{graphicx}
\journalname{Physics of the Dark Universe}
\begin{document}

%\title{Radial oscillations of anisotropic stars within the vanishing complexity factor formalism}
\title{	
Exploring the Structural Properties of Anisotropic Dark Matter-Admixed Quark Stars}	
%\title{Structural Properties of Anisotropic Stars in the Two-Fluid Formalism}
%\thanksref{t1}\right) 


\author{
        Grigoris Panotopoulos \thanksref{e1,addr1}
        \orcidlink{0000-0002-7647-4072}
        \and
        \'Angel Rinc{\'o}n \orcidlink{0000-0001-8069-9162} \thanksref{e2,addr2}
        \and
        Il\'\i dio Lopes 
        \orcidlink{0000-0002-5011-9195}
        \thanksref{e3,addr3}
        \and
%        \and
%        \'Angel Rinc{\'o}n\thanksref{e3,addr2,addr3}
        %etc.
}


\thankstext{e1}{e-mail: 
\href{mailto:grigorios.panotopoulos@ufrontera.cl}
{\nolinkurl{grigorios.panotopoulos@ufrontera.cl}}
}

\thankstext{e2}{e-mail: 
\href{mailto:angel.rincon@ua.es}
{\nolinkurl{angel.rincon@ua.es}}
}

\thankstext{e3}{e-mail: 
\href{mailto:ilidio.lopes@tecnico.ulisboa.pt}
{\nolinkurl{ilidio.lopes@tecnico.ulisboa.pt}}
}


\institute{
Departamento de Ciencias F{\'i}sicas, Universidad de la Frontera, Casilla 54-D, 4811186 Temuco, Chile. \label{addr1}
           \and
Departamento de Física Aplicada, Universidad de Alicante, Campus de San Vicente del Raspeig, E-03690 Alicante, Spain. \label{addr2}
           \and
           Centro de Astrof{\'i}sica e Gravita{\c c}{\~a}o, Departamento de F{\'i}sica, Instituto Superior T{\'e}cnico-IST, Universidade de Lisboa-UL, Av. Rovisco Pais, 1049-001 Lisboa, Portugal. 
           \label{addr3}
           %\and
           %xxxxxx
           %\label{addr3}
}

\date{Received: date / Accepted: date}

\maketitle

\begin{abstract}
We investigate anisotropic compact stars comprising two non-interacting fluids: quark matter and condensed dark matter. 
Using the MIT Bag model equation of state for quark matter and Bose-Einstein Condensate equation of state for dark matter, we numerically compute interior solutions for those two-fluid component spherical configurations. Varying the initial central density ratio of dark matter to quark matter, we examine how different proportions of these components influence the mass-radius profile, the factor of compactness as well as the quark mass fraction. Recent studies suggest that quark matter may exist in the cores of massive neutron stars, significantly affecting their structure and stability. We calculate the factor of compactness for both negative and positive anisotropy cases explored in this article. Our findings demonstrate that dark matter-admixed quark stars are more compact yet less massive compared to pure quark matter stars, aligning with recent theoretical predictions and gravitational wave observations.
\end{abstract}

%%%%%%%%%%%%%%%%%%%%%%%%
\section{Introduction}
\label{sec-intro}
%%%%%%%%%%%%%%%%%%%%%%%

\smallskip\noindent 

The study of compact astrophysical objects, including black holes, neutron stars and white dwarfs, has been fundamental to relativistic and gravitational astrophysics since Karl Schwarzschild's seminal 1916 solution to Einstein's field equations. His solution characterized the spacetime geometry surrounding a static, spherically symmetric mass, revolutionizing our grasp of black holes \cite{1916SPAW.......189S}. This breakthrough catalyzed research into the internal composition and behavior of other compact objects, spawning diverse analytical and numerical models — initially for isotropic stellar systems and later extending to anisotropic configurations.

\smallskip\noindent 
Neutron stars, formed during the final stages of stellar evolution, exhibit extreme internal densities that distinguish them from conventional matter, necessitating their study within the framework of Einstein's General Relativity (GR) \cite{Einstein:1915ca,Shapiro:1983du,Psaltis:2008bb,Lorimer:2008se,Zorotovic:2019uzl}. Understanding these compact objects requires a multidisciplinary approach integrating nuclear physics, astrophysics, and gravitational physics. As the densest known objects after black holes, neutron stars recreate physical conditions unattainable in terrestrial laboratories, functioning as natural cosmic laboratories \cite{2020NatPh..16..907A}. 
Recent research suggests that massive neutron stars may contain quark-matter cores  
{\cite{Alford:2006vz,Weissenborn:2011qu,Orsaria:2019ftf,Baym:2017whm},
challenging previous assumptions about their composition \cite{1999PhRvC..60a5802P}.
Further investigations explore quark stars under various conditions, including within modified gravity frameworks like f(R,T) gravity 
\cite{2025ChPhC..49a5102B} and Rainbow Gravity
\cite{2024PDU....4601610T},
which yield specific mass-radius profiles and observational constraints.  
 This finding has significant implications for neutron star phenomenology and merger dynamics. Additionally, theoretical models of strange quark stars, which may represent the fundamental state of hadronic matter, offer further insights into the properties of dense matter \cite{Alcock:1986hz,Alcock:1988re,Madsen:1998uh,Weber:2004kj,Yue:2006it,Leahy:2007we}.

\smallskip\noindent 
Quark matter, a proposed type of matter made up of free quarks, might exist at the cores of very dense objects like neutron stars. In extreme heat and pressure, neutrons and protons in these stars could break apart into quarks, creating a state of matter not found on Earth. Physicists  have intensely studied the idea of stable quark matter, especially strange quark matter, in both theory and observation \cite{2018PhRvL.120v2001H,2015A&A...577A..40B}. Recent research has challenged the long-held assumption that quark matter must contain strange quarks to be stable, suggesting that up and down quarks alone might form a stable configuration under certain conditions \cite{2018PhRvL.120v2001H,2023NatCo..14.8451A}.
The potential existence of quark matter cores in massive neutron stars has significant implications for our understanding of nuclear physics, the behavior of matter under extreme conditions, and the properties of neutron stars themselves \cite{2015A&A...577A..40B}. 

\smallskip\noindent 
Strong support has emerged from recent research for the existence of quark matter cores in neutron stars, notably in the most massive observed instances \cite{2023NatCo..14.8451A}.
These findings offer crucial insights into the phase diagram of Quantum Chromodynamics (QCD) and the nature of matter at supranuclear densities\cite{2020NatPh..16..907A}.
  
\smallskip\noindent  
Recent studies have highlighted a strong link between dark matter, compact objects, and the key role of dark energy in the evolution of the universe \cite{2018RvMP...90d5002B, 2008ARA&A..46..385F}. Dark matter, which makes up about 27\% of the Universe’s mass-energy, remains mysterious and is only observed through its gravitational effects on galaxies and large-scale structures \cite{2017FrPhy..12l1201Y}. In contrast, dark energy, accounting for roughly 68\% of the Universe’s energy density, is thought to drive the current accelerating expansion of the Universe \cite{2020A&A...641A...6P}. 

\smallskip\noindent  
The true nature of dark matter remains unknown, but theoretical models offer intriguing possibilities. Dark matter is thought to be made up of particles beyond the Standard Model, with leading candidates including Weakly Interacting Massive Particles (WIMPs), axions, and other exotic particles \cite{2010ARA&A..48..495F,2012RAA....12.1107T, 2015PhR...555....1B,2018RPPh...81f6201R, 2019Univ....5..213P,2021A&A...651A.101L}. These studies aim to explore matter under extreme densities and may provide insights into the nature of dark matter.  

\smallskip\noindent  
Theoretical models also suggest that dark matter particles could interact with stars, including compact objects, producing observable effects \cite{2022panu.confE.313S, 2023PhRvD.108h3028L, 2024PhRvD.109d3043B, 2023ApJ...958...49S, 2023ApJ...953..115G, 2020PhRvD.102f3028I,Leung:2011zz,Rezaei:2016zje,Das:2020vng,Mariani:2023wtv}. For instance, recent work by
Liu et al. (2025) \cite{2025arXiv250104382L} explored the effects of bosonic self-interacting dark matter on quark star properties using the confined-isospin-density-dependent-mass model, examining potential implications for observed peculiar objects. Alternativelly, WIMPs might accumulate inside neutron stars, annihilate, and heat their cores, impacting their thermal evolution \cite{2008PhRvD..77b3006K}. Similarly, axions produced in main-sequence and neutron stars could change their structure and emission properties \cite{2016PhRvD..93f5044S, 2023ApJ...943...95S,2021PhRvD.104b3008L,2023JCAP...11..091N}. Primordial black holes, another dark matter candidate, could also be captured by neutron stars, potentially causing mergers that produce detectable gravitational waves \cite{Capela:2013yf}. These possibilities highlight the need for further research and observations to test these ideas and improve our understanding of dark matter.

\smallskip\noindent 
Current and upcoming space missions like Euclid \cite{2020A&A...642A.191E} and the Nancy Grace Roman Space Telescope \cite{2015arXiv150303757S}, along with ground-based surveys such as DESI \cite{2016arXiv161100036D}, are expected to provide a wealth of data that could significantly deepen our understanding of dark matter. These missions aim to uncover the properties of dark matter, shedding light on its nature and its role in shaping the structure and evolution of the universe \cite{2016RPPh...79i6901W}.  

\smallskip\noindent 
Research into dark matter is further supported by gravitational wave detections from compact object mergers, observed by facilities such as LIGO \cite{2016PhRvL.116f1102A} and Virgo \cite{2005AIPC..794..307A}, as well as future projects like LISA \cite{2017arXiv170200786A} and the Einstein Telescope \cite{2010CQGra..27s4002P,2020JCAP...03..050M}. These observations offer a unique way to explore the behaviour of matter under extreme conditions and may help reveal how dark matter interacts with compact objects. 

\smallskip\noindent 
Additionally, X-ray observatories like Chandra \cite{2009ApJ...697.1071A}, XMM -Newton \cite{2001A&A...365L...1J}, and NICER \cite{2014SPIE.9144E..20A} have provided valuable insights into neutron stars and other compact objects, which may interact with dark matter. Future missions such as eXTP \cite{2019SCPMA..6229502Z}, STROBE-X \cite{2019arXiv190303035R}, Lynx \cite{2018arXiv180909642T}, and NewAthena aim to revolutionise studies of these environments. By probing the most energetic phenomena and stellar oscillations, these missions could improve our understanding of the conditions under which dark matter may reveal itself \cite{2013arXiv1306.2307N}.

\smallskip\noindent 
In this study, we investigate anisotropic stars comprising quark and dark matter, employing an MIT Bag equation of state \cite{2024NuPhA104222796P} for quark matter and a barotropic EoS for dark matter. By adjusting the initial central density ratio of dark matter to quark matter, we generate diverse solutions to examine how varying proportions of these components influence the mass-radius profile. Recent studies suggest that quark matter may exist in the cores of massive neutron stars, significantly affecting their structure and stability, as well as their spacetime curvature properties, particularly when considering the influence of dark matter \cite{2025arXiv250111435A}.
 We calculate the compactness factor for the two primary cases and corresponding sub-cases explored in this paper. Additionally, we consider the potential impact of anisotropic pressure on the internal structure of these compact stars, which could alter their tidal deformability and gravitational wave signatures during mergers. For previous similar works see e.g. \cite{Grammenos}, where the authors studied objects made of anisotropic quark matter mixed with dark energy (rather than dark matter), and \cite{Maurya} in which the authors studied electrically charged compact stars with dark matter induced anisotropy. In the first publication an exact analytic solution was obtained, while the anisotropic factor of quark matter was found to be always positive. In the second work the approach of gravitational decoupling was employed to solve the structure equations.

\smallskip\noindent 
The paper is structured as follows: After this introduction, Section \ref{sec-AM} discuss the motivation for 
anisotropic matter in compact stars,
Section \ref{sec-HE}, presents the hydrostatic equilibrium framework for anisotropic stars in the two-fluid formalism, including the modified Tolman-Oppenheimer-Volkoff equations and boundary conditions. Section \ref{sec-EOS} details the equations of state and anisotropy factors for both quark matter (using the MIT Bag model) and dark matter (using a polytropic equation). Section \ref{sec-NS} examines numerical solutions for anisotropic compact stars with varying central densities, analysing both negative and positive anisotropy components. Section \ref{sec-Conclusion} concludes with a summary of our findings on anisotropic compact stars containing both quark and dark matter, with particular emphasis on how their structural properties differ from pure quark matter stars. Throughout the manuscript, we employ geometric units \( G = 1 = c \) and adopt the mostly posotive metric signature  \( \{-,+,+,+\} \). The results demonstrate how different proportions of quark and dark matter influence stellar properties such as mass, radius and compactness.

%%%%%%%%%%%%%%%%%%%%%%%%%%%%%%%%%%%%%%%%%%%%%%%
\section{Anisotropic matter and compact stars}
\label{sec-AM}
%%%%%%%%%%%%%%%%%%%%%%%%%%%%%%%%%%%%%%%%%%%%%%%

In the diverse range of cosmological models examined, the inherent anisotropic properties of matter play a fundamental role, profoundly influencing the behavior and structure of celestial bodies, particularly compact stars such as neutron stars. Anisotropic fluids are crucial to understanding various astrophysical phenomena, including compact objects and potentially dark matter. Recent theoretical work suggests that dark matter may exhibit local anisotropy, which could markedly affect its behaviour and detection. Cadogan and Poisson \cite{2024GReGr..56..118C} present a comprehensive investigation of anisotropic fluids in both Newtonian and relativistic gravity, examining the intricacies of anisotropic matter in astrophysical contexts. Within the framework of General Relativity, anisotropic fluids have been extensively studied, with particular emphasis on their distinctive properties in stellar models and their effects on the structure and behavior of compact objects \cite{2002JMP....43.4889H,2008PhRvD..77b7502H}. Furthermore, research investigating the impact of anisotropy on the mass-radius relationship of stars has yielded remarkable findings, further emphasising the significance of this property in astrophysics \cite{2003RSPSA.459..393M,2022NewAR..9501662K}. This line of inquiry has expanded our understanding of stellar structure and evolution, and remains an active area of research with implications for diverse astrophysical phenomena.

\smallskip\noindent 
Anisotropic stellar models provide a more sophisticated representation of compact objects, as numerous factors can induce pressure anisotropies within stars. These include intense magnetic fields, superfluid states, and exotic matter compositions. Recent studies have demonstrated that anisotropy significantly influences stellar structure and evolution \cite{2024PhRvD.110d3041B}. Indeed, research suggests that anisotropic fluids in neutron stars could substantially alter their mass-radius relationships and affect their stability. However, incorporating anisotropy introduces additional complexity, necessitating an extra degree of freedom to fully resolve the equations. This complexity has spawned innovative approaches in modeling, such as the utilisation of gravitational decoupling techniques to construct physically viable solutions for compact stars with dark matter haloes \cite{1997PhR...286...53H}.

\smallskip\noindent 

The possibility of both positive and negative anisotropy inside stars has been thoroughly investigated in recent years, with compelling evidence supporting both scenarios. Positive anisotropy, where the tangential pressure exceeds the radial pressure, can arise from various physical mechanisms such as strong magnetic fields, rotation, or the presence of exotic matter \cite{2024PhRvD.109d3025B}. This type of anisotropy tends to increase the maximum mass and radius of neutron stars, potentially explaining observations of unusually massive compact objects. Conversely, negative anisotropy, where radial pressure dominates, can result from phase transitions or density-dependent interactions within the stellar core \cite{2024PhRvD.109d3025B}. Recent studies have revealed that negative anisotropy can lead to more compact stellar configurations, affecting properties such as the tidal deformability and gravitational wave signatures \cite{2024JCAP...03..054M}. The presence and magnitude of anisotropy significantly influence stellar structure and evolution, with profound implications for our understanding of extreme matter states and gravitational wave astronomy \cite{2024PhRvD.109d3025B}.

\smallskip\noindent 

This study explores stars containing quark and dark matter, offering valuable insights into dark matter's role in pure quark stars within astrophysics. We examine the possibility of stars with exotic matter types. Our research investigates various scenarios, proposing alternative explanations for compact star and black hole observations, and their implications for neutron stars \cite{2019JCAP...07..012N,2021JCAP...02..045S,2012JCAP...03..037Y,2014arXiv1412.7323T}. The interaction between dark matter and these exotic states \cite{2024PhRvD.109d3043B} could produce distinctive gravitational wave signatures during stellar mergers \cite{2019JCAP...07..012N,2024MNRAS.527.5192M}. These discoveries underscore the necessity for further research into these complex systems to better understand the universe's most enigmatic objects.

%%%%%%%%%%%%%%%%%%%%%%%%%%%%%%%%%%%%%%%%%%%%%%%%%%%%%%%%
\section{Interior solutions in the two-fluid formalism}
\label{sec-HE}
%%%%%%%%%%%%%%%%%%%%%%%%%%%%%%%%%%%%%%%%%%%%%%%%%%%%%%%%

Let us assume a static, spherically symmetric object, described using the usual Schwarzschild-like coordinates: \(x^0 = t\), \(x^1 = r\), \(x^2 = \theta\), and \(x^3 = \phi\). The line element for geometries of non-rotating stars in Schwarzschild coordinates is given by:
\begin{equation}
    ds^2 = -e^{\nu} dt^2 + e^{\lambda} dr^2 + r^2 (d\theta^2 + \sin^2\theta\, d\phi^2), \label{metric}
\end{equation}
where $0 \leq r \leq R$ and $R$ is the stellar radius. The mass function $m(r)$ is defined by:
\begin{equation}
    e^{\lambda} = \frac{1}{1 - \frac{2m}{r}}.
\end{equation}
The Tolman-Oppenheimer-Volkoff (TOV) equations \cite{Tolman:1939jz,Oppenheimer:1939ne} for a two-fluid system are the following \cite{Sandin:2008db,Ciarcelluti:2010ji}:
\begin{align}
 m'(r) &= 4\pi r^2 \rho(r), \\
 \nu'(r) &= 2 g(r), \\
 p_{\rm QM}'(r) &= -[\rho_{\rm QM}(r) + p_{\rm QM}(r)] \, g(r) + 2\frac{\Delta_{\rm QM}(r)}{r}, \\
 p_{\rm DM}'(r) &= -[\rho_{\rm DM}(r) + p_{\rm DM}(r)]\,  g(r) + 2\frac{\Delta_{\rm DM}(r)}{r}, \\
 g(r) &=
\frac{m(r) + 4\pi r^3 p(r)}{r^2 \left(1 - \frac{2m(r)}{r}\right)}, 
\end{align}
where a prime denotes differentiation concerning \(r\), and where \(p_{\rm QM}\) and \(\rho_{\rm QM}\) are the pressure and the energy density for quark matter, respectively, while \(p_{\rm DM}\) and \(\rho_{\rm DM}\) are the corresponding quantities for dark matter. The total pressure and density are, respectively
\begin{align}
 \rho(r) &\equiv \rho_{\rm QM}(r) + \rho_{\rm DM}(r), \\
  p(r)   &\equiv p_{\rm QM}(r) + p_{\rm DM}(r),
\end{align}
The function \(\nu(r)\) may be computed with the help of
\begin{align}
\nu(r) &= \ln \left(1 - \frac{2M}{R}\right) - 2 \int_r^R  g(x) \, dx,
\end{align}
where $M$ is the mass of the star. The equations for $\{ m', p'_{\rm QM}, p'_{\rm DM} \}$ are then integrated with the help of the conditions at the center of the star, $r \rightarrow 0$, i.e., 
\begin{align}
m(0)  &= 0, \\
p_{\rm QM}(0) &= p_{\rm c,QM}, \\
 p_{\rm DM}(0) &= p_{\rm c,DM},
\end{align}
where \( p_{\rm cQM} \) is the central pressure for quark matter and \( p_{\rm cDM} \) for dark matter. 
The latter were the conditions on the center, however, it is also required to impose appropriate conditions at the surface. Thus, the following matching conditions must be satisfied at the surface of the object, $r \rightarrow R$: 
\begin{align}
p(R) &= 0, \\ 
m(R) &= M,
\end{align}
taking into account that the exterior vacuum solution is given by the Schwarzschild geometry \cite{1916SPAW.......189S}.
Equivalently, the first and second fundamental forms across that surface confirm the last comments. In particular, recall that
\begin{subequations}
\begin{eqnarray}
e^{\nu(r)} \Bigl|_{r=R} &=& 1-\frac{2M}{R},
\label{enusigma}
\\
e^{\lambda(r)} \Bigl|_{r=R} &=& \Bigg(1-\frac{2M}{R}\Bigg)^{-1},
\label{elambdasigma}
\\
p(r) \Bigl|_{r=R} &=& 0, 
\label{PQ}
\end{eqnarray}
\end{subequations}
and such conditions are precisely the necessary and sufficient conditions for a smooth matching of the two metric potentials (\ref{enusigma}) and (\ref{elambdasigma}) at the surface $r=R$. These boundary conditions ensure that the interior solution, described by the metric (\ref{metric}), seamlessly joins the exterior vacuum solution given by the Schwarzschild space-time at the stellar surface.

%%%%%%%%%%%%%%%%%%%%%%%%%%%%%%%%%%%%%%%%%%%%%%%%%%%%%%%%%%%%%%%%%%%
\section{Equations-of-state and anisotropic matter in the two-fluid formalism}
\label{sec-EOS}
%%%%%%%%%%%%%%%%%%%%%%%%%%%%%%%%%%%%%%%%%%%%%%%%%%%%%%%%%%%%%%%%%%%
\smallskip\noindent 

Herein, we shall present the theory and equations for examining compact stars comprised of two distinct fluids: quark matter and dark matter. We shall investigate two specific cases: i) negative anisotropic factor and ii) positive anisotropic factor, both pertaining to quark matter. Subsequently, we shall discuss the equation-of-state (EoS) describing both quark and dark matter, alongside the specific form of the anisotropic factor for each fluid.

%%%%%%%%%%%%%%%%%%%%%%%%%%%%%%%%%%%%%%%%%%%%%%%%%%%%%%%
\subsection{Assumptions of EoS and anisotropic factor}
%%%%%%%%%%%%%%%%%%%%%%%%%%%%%%%%%%%%%%%%%%%%%%%%%%%%%%%

\smallskip\noindent 

We should mention that our aim is to integrate the TOV equations to obtain the corresponding numerical profiles of the functions. We then have to assume the equation of state of the matter content and, if the fluid is anisotropic, some concrete form of the anisotropic factor.
In the two-fluid formalism, we take advantage of the fact that there is no direct interaction between the two fluid components. The latter means that each fluid can be properly described by its own equation of state.
Let us start by considering one equation of state for the quark matter content modeled in this case by the MIT Bag model EoS\cite{Chodos:1974je,Chodos:1974pn}, a framework also utilized in studies searching for other compact hadronic states like pentaquarks\cite{2025arXiv250112727L}, i.e.,
\begin{align}
    p_{\rm QM}(r) &= \frac{1}{3}\Bigl( \rho_{\rm QM}(r) - 4B \Bigl),
\end{align}
where the bag constant $B$ has been constrained to take values in the range \cite{Aziz:2019rgf}
\begin{align}
   & 41.58 \: \frac{\text{MeV}}{\text{fm}^3} \leq B \leq 
    319.31 \: \frac{\text{MeV}}{\text{fm}^3}.
\end{align}
It is related to the value of energy density on the surface $\rho_{\rm sQM}$, after imposing $p(R)=0$ to generate $\rho_{\rm QM}(R) \equiv \rho_{\rm sQM} = 4B$. 
In addition, we are considering this fluid as anisotropic, so, becomes essential to include the anisotropic part. For simplicity, following \cite{Arbanil:2021ahh} we introduce 
\begin{align}
    \Delta_{\rm QM}(r) &= \kappa_1 \: p_{\rm QM}(r) \: \frac{2 m(r)}{r}.
\end{align}
where $\kappa_1$ is a dimensionless constant parameter, while both $p_{\rm QM}(r)$ and $m(r)$ have the usual meaning.
Subsequently, the dark matter content, reinterpreted as a Bose-Einstein Condensate, may be effectively described via a polytropic EoS, namely:
\begin{align}
    p_{\rm DM}(r) &= K \rho_{\rm DM}(r)^2
\end{align}
with index $n=1$, where $K$ is a constant of proportionality with the right dimensions and $\rho_{\rm DM}$ is the energy density of dark matter. Dark matter was reinterpreted as a Bose-Einstein Condensate some time ago \cite{Boehmer:2007um,Pires:2012yr,Souza:2014sgy}, which was able to address the cusp/core problem \cite{Harko:2011xw}. 
The single wave function describing the Bose-Einstein condensation \cite{Bose1924Plancks,einstein1924,einstein1925} satisfies the Gross-Pitaevskii equation \cite{1999RvMP...71..463D,Pethick2002}. The presence of the non-linear term, interpreted as enthalpy, leads to a polytropic equation of state (EoS), as highlighted e.g. in \cite{Li:2012sg,Chavanis:2017loo}. 

\smallskip\noindent 

We comment in passing that the precise form of the EoS depends on the form of the self-interaction scalar potential. However, in the low density limit, the EoS always takes the form of the above polytrope. Consider for instance the quartic potential, $V(\phi) \sim \lambda \phi^4$, with $\lambda$ being a dimensionless coupling constant. It was shown in \cite{Colpi:1986ye} that the pressure as a function of the energy density is given by $p=(\rho_0/3) (\sqrt{1+\rho/\rho_0}-1)^2$ where the constant $\rho_0$ is defined by $\rho_0 \equiv m^4/(3 \lambda)$ with $m$ being the mass of the scalar field. It is easy to see that when $\rho \ll \rho_0$ the EoS takes the approximate form $p \propto \rho^2$ \cite{Maselli:2017vfi, Moraes:2021lhh}, while in the high density regime, $\rho \gg \rho_0$, it takes the usual form $p \approx \rho/3$ for ultra-relativistic matter \cite{Maselli:2017vfi}.

\smallskip\noindent 

Identically to the quark matter case, we are interested in investigating anisotropic fluids, so, we assume an anisotropic contribution similar to the previous case, but adapting the parameters, namely:
\begin{align}
    \Delta_{\rm DM}(r) &= \kappa_2 \: p_{\rm DM}(r) \: \frac{2 m(r)}{r}. 
\end{align}
with a different dimensionless coupling $\kappa_2$.

\smallskip\noindent 

Before we proceed to present and discuss the numerical results, a comment is in order here. In the present study we have adopted the 2-fluid formalism assuming that there is no direct interaction between quarks and DM. This possibility may be realized for instance in one of the simplest extensions of the Standard Model of Particle Physics, namely the real gauge-singlet scalar with a discrete $Z_2$ symmetry, discussed e.g. in \cite{Lerner:2009xg,Biswas:2011td}. In this model the total Lagrangian density is given by $\mathcal{L}=\mathcal{L}_{SM} + \mathcal{L}_{DM}$, where the DM part is given by $\mathcal{L}_{DM} = (1/2) (\partial S)^2 - V(S) - g h^2 S^2$ \cite{Lerner:2009xg,Biswas:2011td}, where $g$ is a dimensionless coupling constant between the Higgs boson and the gauge-singlet scalar field $S$. Since the trilinear interactions of the form $hSS, Shh$ are not allowed by the symmetries, the elastic scattering process $DM \: N \rightarrow DM \: N$, with $N$ being the nucleon, does not take place via the Higgs exchange, see e.g. Fig. 1 of \cite{Andreas:2008xy}. This avoids the discussion about constraints on the DM-nucleon cross-section. In addition to that, the usual discussion about DM capture from the halo and its thermalization inside stars \cite{Gould:1987ww,Kouvaris:2010vv,Kouvaris:2015nsa,Panotopoulos:2020kuo} is not relevant in the present study.

%%%%%%%%%%%%%%%%%%%%%%%%%

%%%%%%%%%%%%%%%%%%%%%%%%%%%%%%
\section{Numerical solutions}
\label{sec-NS}
%%%%%%%%%%%%%%%%%%%%%%%%%%%%%%

\smallskip\noindent 

In this work we have studied anisotropic stars consisting of two different fluids, one quark matter and the other dark matter, in the framework of General Relativity, taking advantage of some well-motivated EoS and anisotropic factors. We have numerically computed interior solutions of (realistic) spherical configurations of anisotropic matter (quark and dark matter) varying the central densities of the fluid components at the center of the star. 

\smallskip\noindent 
For convenience, we have introduced the dimensionless factor $f$ defined by:
\begin{equation}
f \equiv \frac{\rho_{c,\text{DM}}}{\rho_{c,\text{DM}}+\rho_{c,\text{QM}}} ,   
\end{equation}
For a given $f$ the central energy densities are not independent, and we may express one in terms of the other, for instance
\begin{align}
    \rho_{c,\text{DM}} &=  \left(\frac{f}{1-f} \right) \rho_{c,\text{QM}} \equiv \alpha \: \rho_{c,\text{QM}},
\end{align}
or in other words, the central density of dark matter is proportional to the central density of quark matter, and we have defined a new dimensionless factor
\begin{equation}
\alpha \equiv \frac{f}{1-f}.
\end{equation}
As the anisotropic factor of dark matter condensate is always negative \cite{Moraes:2021lhh}, in the present study we have considered two cases, namely: a) anisotropic quark matter where $\Delta_{\text{QM}}(r) < 0$ (case I) assuming
\begin{align}
f &= 0.58, \; 0.60, \; 0.62 \text{ (or equivalently)}
\\
\alpha &\approx 1.38 , \; 1.50 , \; 1.63
\end{align}
and b) anisotropic quark matter characterized by a positive anisotropic factor, $\Delta_{QM}(r) > 0$ (case II), assuming the same values of case I for $f$ (or equivalently $\alpha$).
% %
% \begin{align}
% f &= 0.59, \; 0.60, \; 0.62 \text{ (or equivalently)}
% \\
% \alpha &\approx 1.43, \; 1.50, \; 1.63. 
% \end{align}
% %
Our main results are discussed in the next subsections.
In what follows, we shall consider the following numerical values to perform our study
\begin{align}
    B &= 57.64 \: \frac{\text{MeV}}{\text{fm}^3}
    \\
    K &= \frac{0.01}{B}
    \\
    \kappa_2 &= - 0.2
    \\
    \kappa_1 &= \pm 0.1.
\end{align}
As mentioned before, to obtain our numerical results, a condition must be imposed at the center of the star. Within the two-fluid formalism, we relate the central densities of the two fluids using the parameter $\alpha$, which proves more convenient than $f$ for characterizing variations in the central density. We consider three representative cases where the central density of dark matter exceeds that of quark matter by i) 38$\%$, ii) 50$\%$, and iii) 63$\%$. 
These percentages provide a simple and effective way of understanding how changes in the central density of dark matter affect the structure and evolution of the star. By exploring a wide range of density contrasts, we can study how sensitive different mixed star configurations are to the presence of dark matter.

Last but not least, it is also worth mentioning at this point that the anisotropic factors $\Delta_{QM}(r)$ and $\Delta_{DM}(r)$ are proportional to $\kappa_1$ and $\kappa_2$, respectively, and these factors encode the strength of the anisotropy over each fluid. Thus, we examine low values of these parameters in order to consider fluids that are slightly anisotropic.

%{\bf 
Finally, we rely on the criterion based on the relativistic adiabatic index defined by
\begin{equation}
\Gamma = \frac{d p}{d \rho} \: \left( 1 + \frac{\rho}{p} \right)
\end{equation}
to check whether or not the configurations considered here are stable. The condition $\Gamma \geq \Gamma_{cr}$ ensures stability, where the critical value is computed by $\Gamma_{cr}=4/3 + (19/21) (M/R)$ \cite{Moustakidis:2016ndw}.
%}

%%%%%%%%%%%

%%%%%%%%%%%%%%%%%%%%%%%%%%%%%%%%%%%%%%%%%%%%%%%%%%%%%%%%%%%%
\subsection{{\bf Case I}: Negative anisotropy of quark matter}
%%%%%%%%%%%%%%%%%%%%%%%%%%%%%%%%%%%%%%%%%%%%%%%%%%%%%%%%%%%%

\smallskip\noindent 

Our analysis examines the compactness factor, $C$, (defined as usual, i.e., $M/R$) versus stellar mass, $M$, (in solar masses)  across three scenarios varying the $\alpha$ or $f$ factors. 
As shown in Figure  \ref{fig:1} (top panels), when the central density of dark matter increases relative to quark matter's central density, the compactness factor exhibits a slight leftward shift while maintaining its upward trend \cite{2020NatPh..16..907A}.
The M-R profiles demonstrate distinct characteristics through four curves. Three solid curves represent dark matter-admixed quark stars with different initial (central) conditions, while a dashed black line indicates objects made of pure quark matter. The two-fluid configurations produce more compact yet less massive objects compared to pure quark stars, aligning with recent theoretical predictions \cite{2024MNRAS.527.5192M}.
This demonstrates dark matter's significant influence on stellar properties when quark matter is present \cite{2024PhRvD.109d3043B}, consistent with gravitational wave observations \cite{2019JCAP...07..012N}. As the central density ratio ($\alpha$) of dark matter to quark matter increases, the mass-radius profile evolves systematically.
The object becomes more compact, while its highest mass slightly decreases, supporting recent theoretical frameworks \cite{2024NuPhA104222796P}.
Similarly, investigations within alternative frameworks like Rastall gravity also analyze how varying model parameters influence the mass-radius relation and stability of quark stars in comparison with recent astrophysical observations, as explored in recent studies \cite{2024ChJPh..90..422T,2025CQGra..42b5008B}.
For non-zero but modest $\alpha$ values, we obtain physically viable solutions aligned with conventional compact star parameters, particularly matching observed mass-radius relationships in recent surveys 
\cite{2023ApJ...958...49S,2023EPJC...83.1065R}. Finally, since the spherical configurations studied here are made of two fluid components, it is interesting to see how the quark mass fraction changes as we vary the central energy density of quark matter. According to the bottom right panel, the mass fraction increases with the central density, it reaches a maximum value of $\sim 65 \%$ around $\rho_{c,QM} \approx 2.75 \rho_s$ (green curve), and then it monotonically decreases. Its highest value depends on the numerical value of the factor $f$, and as our numerical results indicate, the curves are shifted downwards as $f$ increases.

\smallskip\noindent 

Figure \ref{fig:1} presents the mass-to-radius profiles, compactness factor, and quark mass fraction for anisotropic dark matter-admixed quark stars in the case of negative quark matter anisotropy. Notably, the inclusion of up-to-date astrophysical constraints, such as the precisely measured masses of PSR~J1614$-$2230 \cite{Demorest},
PSR~J0348+0432 \cite{Antoniadis:2013pzd}, and PSR~J0740+6620 \cite{
Miller:2019cac,Riley:2021pdl}
is essential for evaluating the viability of the theoretical models shown. The yellow and brown horizontal bands, as well as shaded regions representing the latest NICER measurements for PSR~J0030+0451 and PSR~J0740+6620 \cite{Miller:2019cac,2019ApJ...887L..21R,Miller:2021qha,Riley:2021pdl}, clearly demonstrate that only certain regions of the parameter space remain consistent with observational data. The dark matter- admixed models generally result in more compact yet less massive configurations relative 
to pure quark stars, often falling below the $2\,M_\odot$ constraint set by pulsar observations. This direct comparison illustrates the impact of increasing dark matter content on the maximum mass and compactness, highlighting the importance of integrating observational constraints in 
the analysis of exotic compact objects.}

%%%%%%%%%%%%%%%%%%%%%%%%%%%%%%%%%%%%%%%%%%%%%%%%%%%%%%%%%%%
\subsection{{\bf Case II}: Positive anisotropy of quark matter}
%%%%%%%%%%%%%%%%%%%%%%%%%%%%%%%%%%%%%%%%%%%%%%%%%%%%%%%%%%%

\smallskip\noindent 

\begin{figure*}[ht!]
\centering
\includegraphics[scale=0.935]{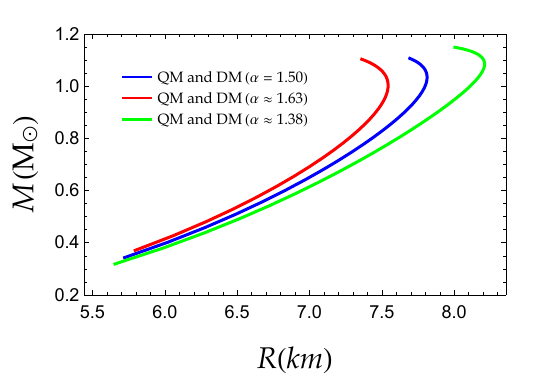} \
\includegraphics[scale=0.935]{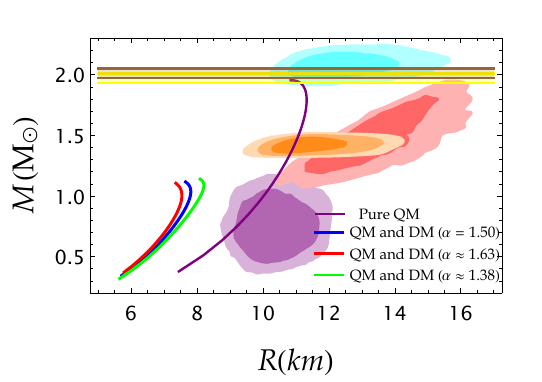} 
\\
\includegraphics[scale=0.935]{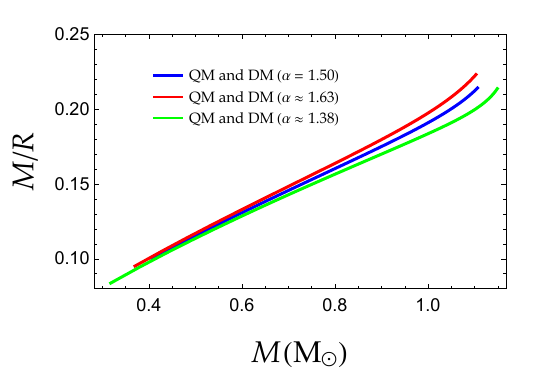} \
\includegraphics[scale=0.935]{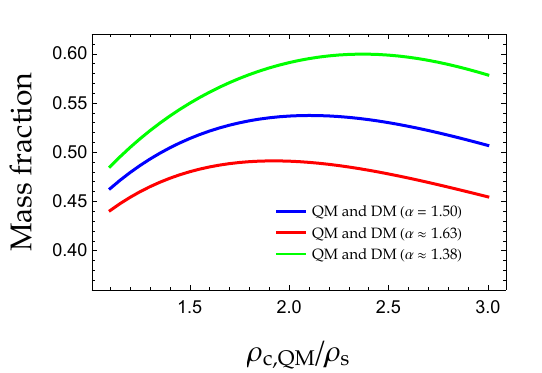} 
 \caption{
Mass-to-radius profiles, factor of compactness and quark mass fraction for {\bf case I} studied here, i.e. for negative $\kappa_1$. 
{\bf Top panels:} Stellar mass in solar masses versus stellar radius in km assuming $f=0.58$ (green), $f=0.60$ (blue) and $f=0.62$ (red). 
For comparison reasons, in the upper right panel we have also shown the mass-to-radius relationship without the inclusion of dark matter, i.e. pure quark matter. 
We have included two horizontal strips around two solar masses. The yellow band corresponds to the massive pulsar J1614-2230, which has a known mass of $M=(1.97\pm0.04)M_{\odot}$ \cite{Demorest}. The brown band corresponds to the massive pulsar J0348+0432, which has a known mass of $M=(2.01\pm0.04)M_{\odot}$ \cite{Antoniadis:2013pzd}.
In addition, there are four regions corresponding to:
a) the light HESS compact object (purple region) \cite{Doroshenko:2022nwp},
b) the pulsar J0740+6620 (cyan region) \cite{Miller:2021qha,Riley:2021pdl}, 
c) the pulsar J0030+0451 (red region), \cite{Miller:2019cac}, 
and 
d) the light HESS compact object (purple region) \cite{Choudhury:2024xbk}. 
The intensity of the color represents 65$\%$, 90$\%$ and 99$\%$ from darker to lighter color.
{\bf{Bottom-Left panel:}} Factor of compactness, $C=M/R$, versus stellar mass (in solar masses) considering $f=0.58,0.60,0.62$.
  {\bf{Bottom-Right panel:}} Quark mass fraction as a function of the (normalized) central energy density of quark matter.
 }
\label{fig:1} 	
\end{figure*}

\begin{figure*}[ht!]
\centering
\includegraphics[scale=0.935]{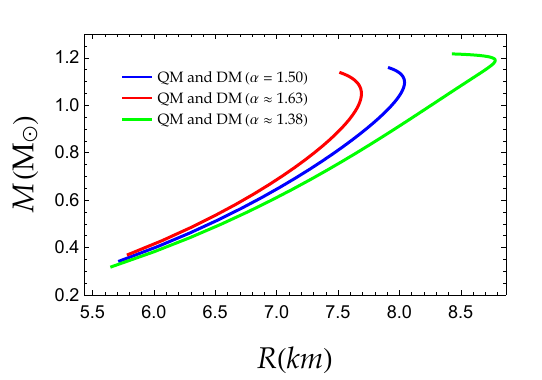} \
\includegraphics[scale=0.935]{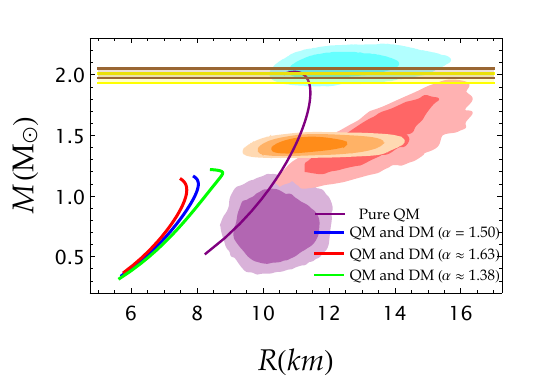} 
\\
\includegraphics[scale=0.935]{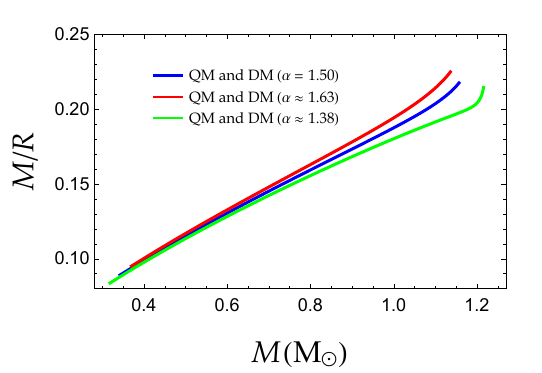} \
\includegraphics[scale=0.935]{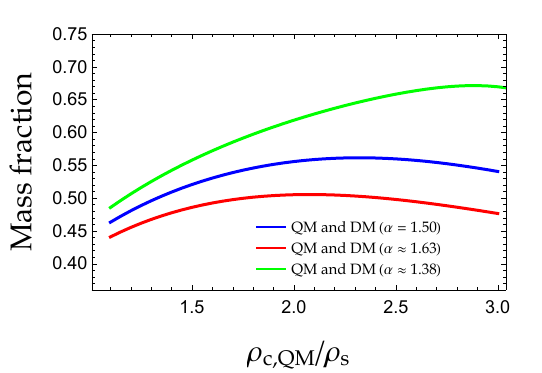} 
 \caption{
Same as in previous figure, but for {\bf case II} studied here, i.e. for positive $\kappa_1$, and for $f=0.58,0.60,0.62$.
}
\label{fig:2} 	
\end{figure*}

When examining positive anisotropy, our analysis of the compactness factor $C$ versus mass $M$ again considers three scenarios with varying $\alpha$ parameters. As illustrated in Fig. \ref{fig:2} (bottom panels), increasing central dark matter density relative to quark matter shifts the compactness factor slightly rightward while maintaining its increasing trend.
The mass-radius profiles reveal distinct characteristics across four curves. Three solid curves represent hybrid quark-dark matter configurations with different central conditions, while a dashed black line indicates pure quark matter stars. The hybrid configurations exhibit less compact yet more massive structures compared to pure quark stars, aligning with recent theoretical work \cite{2006MPLA...21.1965J}.
This demonstrates dark matter's significant influence on stellar properties when quark matter is present \cite{2018PhRvD..97l3007E}, matching gravitational wave observations \cite{2018PhRvL.121p1101A}. As the central density ratio $\alpha$ increases, the mass-radius profile evolves systematically. The object becomes less compact while its maximum mass slightly increases, supporting recent theoretical frameworks \cite{2019PhRvL.122f1102B}. For non-zero but modest $\alpha$ values, we obtain physically viable solutions matching conventional compact star parameters and observed mass-radius relationships \cite{2019ApJ...887L..21R}.
The positive anisotropy component leads to increased maximum mass and radius compared to the isotropic case, becoming more pronounced as anisotropy increases. This allows for more stable configurations at higher central densities, potentially expanding viable quark star models \cite{2020PhRvD.102l4017M}. Additionally, positive anisotropy affects tidal deformability, generally increasing the tidal Love number with implications for gravitational wave signals from binary mergers \cite{2018ApJ...852L..29R}. As far as the quark mass fraction is concerned, it exhibits a behavior very similar to the one shown in Figure \ref{fig:1}.

\smallskip\noindent 

Figure \ref{fig:2} illustrates the structural properties for the case of positive quark matter anisotropy. In this scenario, hybrid stars with increased dark matter content tend to display less compact but more massive profiles, sometimes exceeding the masses of their pure quark matter counterparts for the same central energy density. However, when current astrophysical constraints are overlaid, the physically viable solutions are more stringently limited. The models are required to remain within the allowed mass-radius regions set by multimessenger gravitational wave analyses of GW170817 \cite{
2017PhRvL.119p1101A,
2018PhRvL.121p1101A,
2020NatPh..16..907A} and the latest NICER/XMM-Newton results \cite{
Miller:2021qha,Riley:2021pdl}. These results make it clear that only certain combinations of anisotropy and dark matter fraction are compatible with the observed properties of neutron stars, providing a powerful test for the underlying equation of state and the dark matter hypothesis.

\smallskip\noindent 

\begin{figure*}[ht!]
\centering
\includegraphics[scale=0.722]{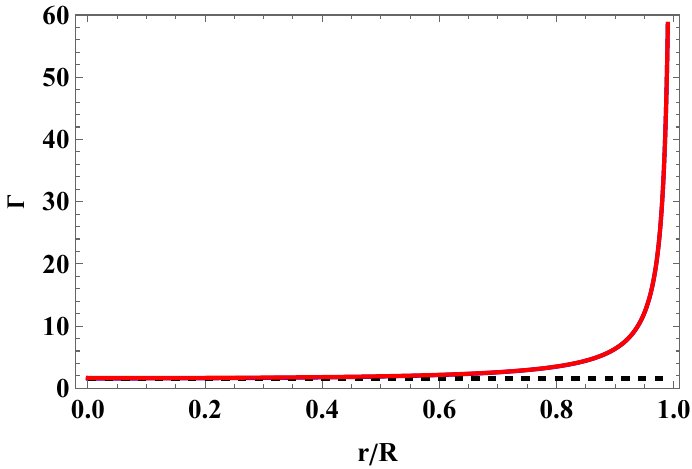} \
\includegraphics[scale=0.722]{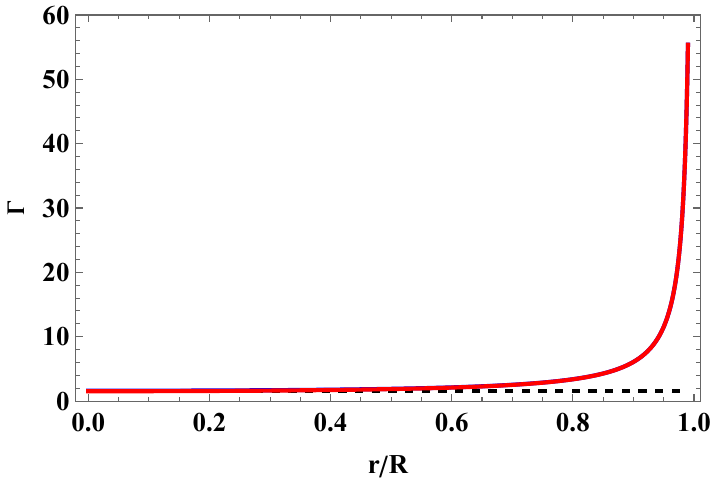} 
\caption{
{\bf{Left panel:}} Relativistic adiabatic index, $\Gamma$, versus normalized radial coordinate, $r/R$, for case I.
{\bf{Right panel:}} Relativistic adiabatic index, $\Gamma$, versus normalized radial coordinate, $r/R$, for case II.
%
%Relativistic adiabatic index, $\Gamma$, versus normalized radial coordinate, $r/R$, for case I (upper panel) and case II (lower panel). 
%
The dashed line corresponds to the critical value $\Gamma_{cr}=4/3 + (19/21) (M/R) $.
}
\label{fig:3} 	
\end{figure*}

In Figure \ref{fig:3} we have shown the relativistic adiabatic index as a function of the radial coordinate for both cases considered in this work. The configurations correspond to the highest mass supported by the quark EoS without the presence of DM. We see that the stability criterion based on $\Gamma$ is met, and therefore the configurations are stable. This guarantees that all other configurations must be stable as well, since both the stellar mass and the factor of compactness are lower compared to the point of highest mass. Next, as shown in the Figures, the inclusion of DM reduces significantly the maximum stellar mass, and therefore we conclude that the DM admixed quark stars, too, must be stable. See also \cite{Jetzer:1990xa} for a stability analysis of combined boson-fermion spherically symmetric configurations.

\smallskip\noindent 

Stability of stars may be also discussed via the study of radial oscillations \cite{Chandra1,Chandra2}, and computation of the frequencies $\omega_0, \omega_1,...$. Stability against radial oscillations for a multi-component star must be carefully studied as in \cite{Kain, Caballero}. A general result of those works is that the maximum mass configuration does not necessarily coincides with the last stable configuration, and that an extended branch of stable configurations might arise, as found for hybrid stars in \cite{Pereira,Lugones} and references therein. The $\omega^2 = 0$ separates the stable from the unstable modes. The methods to solve the eigenvalue value problem for the frequencies are well known. Chandrasekhar in \cite{Chandra2} using trial functions found the critical value $\Gamma_{cr}=4/3 + (19 M)/(21 R)$, where the first term corresponds to the Newtonian value, while the second one is the relativistic correction to that.

\smallskip\noindent 

Another quantity of interest that has attracted a lot of attention is the so called dimensionless tidal deformability $\Lambda$ of relativistic stars in binaries. The recent observation of two gravitational wave events, GW170817 \cite{LIGO1} and GW190425 \cite{LIGO2} put a well-defined EoS independent constraint on the deformability as $\Lambda \leq 800$. See for instance \cite{Biswas,Das} for anisotropic compact objects, and \cite{Li} for quark stars. The computation of tidal Love numbers and deformabilities of DM admixed quark stars is under construction in a separate work \cite{progress}.

\smallskip\noindent 

Finally, before we summarize our work a final comment is in order. According to the bottom-right panel of Figs.  \ref{fig:1} and \ref{fig:2}, the DM mass fraction is comparable to the quark mass fraction. This may rise the concern that the DM mass fraction is too big, since it is estimated that it should be around $10^{-14}$. This number, however, is obtained in the scenario in which there is a non-vanishing DM-nucleon cross-section, and DM capture from the halo. As mentioned before, however, in the present study we have in mind a concrete Particle Physics model where there is no DM-nucleon interaction, and so this estimate no longer holds. The DM mass fraction observed in the figures is of the same order of magnitude considered in \cite{Li:2012qf} (parameter $\epsilon$ there), see Fig. 8 of \cite{Li:2012qf}.

\smallskip\noindent 

A comprehensive examination of Figures \ref{fig:1}, \ref{fig:2}, and \ref{fig:3} in the context of modern observational constraints offers critical insights into the interplay between quark matter, dark matter, and stellar anisotropy. The simultaneous inclusion of radio timing, gravitational wave, and X-ray observations -- covering mass, radius, and compactness -- significantly restricts the allowed parameter space for dark matter-admixed quark stars. In particular, Figure \ref{fig:3}  confirms that the most massive stable configurations remain consistent with the adiabatic index stability criterion, but the addition of dark matter generally reduces the maximum mass achievable. By overlaying these observational bands and constraints onto theoretical results, the study not only enhances its relevance but also demonstrates that the presence and properties of dark matter in compact stars may soon be stringently tested or ruled out with forthcoming observational data. This integrative approach paves the way for more refined future studies exploring the microphysical properties of dense matter and the cosmic role of dark matter in astrophysical compact objects.

%%%%%%%%%%%%%%%%%%%%%%%%%%%%%PLOTS%%%%%%%%%%%%%%%%%%%%%%%%%%%%%
  
%%%%%%%%%%%%%

%%%%

\section{Summary and concluding remarks}\label{Conc}
\label{sec-Conclusion}
%%%%%%%%%%%%%%%%%%%%%%%%%%%%%%%%%%%

\smallskip\noindent 

To summarize our work, this study examined the impact of the inclusion of dark matter on structural properties of quark stars, revealing new insights about those less conventional celestial objects' structure and properties. Our investigation into anisotropic stars consisting of both quark and dark matter has revealed significant insights into their structure and properties.

\smallskip\noindent 

The incorporation of dark matter into quark star models has demonstrated a marked influence on stellar characteristics. In the case of negative quark anisotropy, we observed that quark-dark matter configurations resulted in more compact yet less massive objects compared to pure quark stars \cite{2024MNRAS.527.5192M}.
This aligns with recent theoretical predictions and gravitational wave observations, highlighting the significant role of dark matter in shaping stellar properties when quark matter is present
\cite{2019JCAP...07..012N}. Conversely, for positive quark anisotropy, our findings indicate that two-fluid component configurations exhibit less compact yet more massive structures compared to pure quark stars \cite{2024MNRAS.527.5192M}. This supports recent theoretical frameworks and matches observed mass-radius relationships in recent surveys \cite{2023ApJ...958...49S,2023EPJC...83.1065R}. 
The systematic evolution of mass-radius profiles as the central density ratio of dark matter to quark matter increases provides valuable insights into the internal composition of these exotic stars. For non-zero but modest central density ratios, we obtained physically viable solutions that align with conventional compact star parameters\cite{2023ApJ...958...49S,2023EPJC...83.1065R}.
In general, it is worth emphasizing that, regardless of whether the anisotropic quark matter satisfies $\Delta_{QM}(r) > 0$ or $\Delta_{QM}(r) < 0$, the presence of dark matter in a quark matter star significantly reduces its mass-radius profile. In other words, the resulting configuration is a more compact and lighter object, as illustrated in the upper right panels of Figs. \eqref{fig:1} and \eqref{fig:2}. Furthermore, 
the numerical results clearly fall outside the constraint band established by \cite{Demorest}, as these figures also demonstrate.

\smallskip\noindent 

Our results contribute to the growing body of evidence suggesting the presence of quark matter cores in massive neutron stars, which significantly affects their structure and stability \cite{2020NatPh..16..907A}.
This study also underscores the potential of gravitational wave Astronomy in probing the nature of ultra-dense matter and the role of compact objects in the expansion of the Universe
\cite{2016PhRvL.116f1102A}.

\smallskip\noindent 

Future research directions could include investigating the effects of different equations of state for both quark and dark matter, exploring the impact of varying anisotropy factors, and examining the gravitational wave signatures of these hybrid compact objects. Additionally, upcoming space missions and ground-based surveys promise to deliver new data that could further transform our understanding of these enigmatic stellar remnants and their role in the cosmos \cite{2010CQGra..27s4002P}.

\smallskip\noindent 

In the broader context of Astrophysics and Cosmology, this study contributes to our understanding of the fundamental nature of matter under extreme conditions and the potential role of dark matter in compact objects. As we continue to push the boundaries of our knowledge, the interplay between quark matter and dark matter in compact stars remains a fertile ground for future discoveries and insights into the nature of our Universe.

%%%%%%%%%%%%%%%%%%%%%%%%%%%%
\section*{Acknowledgments}
%%%%%%%%%%%%%%%%%%%%%%%%%%%

%{\bf
We wish to thank the anonymous referees for a careful reading of the manuscript as well as for useful comments and suggestions.
%} 
A.~R. acknowledges financial support from Conselleria d'Educació, Cultura, Universitats i Ocupació de la Generalitat Valenciana thorugh Prometeo Project CIPROM /2022/13.
A.~R. is funded by the María Zambrano contract ZAMBRANO 21-25 (Spain) (with funding from NextGenerationEU).
I.~L. thanks the Funda\c c\~ao para a Ci\^encia e Tecnologia (FCT), Portugal, for the financial support to the Center for Astrophysics and Gravitation (CENTRA/IST/ULisboa)  through the Grant Project No.~UIDB/00099/2020. 

%%%%%%%%%%%%%%%%%%%%%%5

%\section*{Data Availability Statement}
%We do not have any additional data to present. All the numerical results are summarized in several tables in the manuscript.

%
% BibTeX users please use
%\nocite{*} % para citar todos los paper, no solo los citados realmente

\bibliographystyle{unsrt}
%%\bibliography{biblio_1.bib}% Produces the bibliography via BibTeX.
% Ilidio
\bibliography{biblioIL2.bib}

%\begin{thebibliography}{}
%\end{thebibliography}

\end{document}